\documentclass[
    ,final            
  ]
  {aipproc}

\layoutstyle{6x9}

\begin{document}

\title{Deeply virtual Compton scattering and generalized parton distributions}


\author{S.~Fazio}{
  address={\sl Dipartimento di Fisica, Universit\`a della Calabria \\
and Istituto Nazionale di Fisica Nucleare,
Gruppo collegato di Cosenza} \centerline{  \sl I-87036 Arcavacata
di Rende, Cosenza, Italy} }

\author{R.~Fiore}{
  address={\sl Dipartimento di Fisica, Universit\`a della Calabria \\
and Istituto Nazionale di Fisica Nucleare,
Gruppo collegato di Cosenza} \centerline{  \sl I-87036 Arcavacata
di Rende, Cosenza, Italy} }

\author{L.L. Jenkovszky}{
  address={Bogolyubov Institute for
Theoretical Physics, Nat. Ac. Sciences, Kiev-143, 03680
Ukraine,}\\ {RMKI (KFKI), POB 49, Budapest 114, H-1525 Hungary }}

\begin{abstract}
We present a comparison of a recently proposed model, which describes the Deeply 
Virtual Compton Scattering amplitude, to the HERA data.
\end{abstract}

\maketitle


Exclusive production of a real photon, a vector meson or 
lepton pairs via deeply virtual scattering, $ep\rightarrow eVp$, 
where $V$ stands generically for produced particles, 
is an interesting tool to investigate the diffractive properties 
of the Pomeron.

The linear Regge-trajectory of the Pomeron is,
\begin{equation}
    \alpha_P(t,Q^2)=\alpha_0(Q^2)+\alpha '(Q^2)t,
\label{trlin}
\end{equation}
where $t$ is the squared four-momentum transferred at the proton
vertex and $Q^2$ is the virtuality of the exchanged photon. The 
standard Regge pole parametrization of the scattering
amplitude is  
\begin{equation}
A(s,t,Q^2)=A_0e^{B(t,Q^2)}\Biggl({s\over{s_0}}\Biggr)^{\alpha_P(t,Q^2)},
\end{equation}
where $s=W^2$ is the squared $\gamma^* p$ centre-of-mass energy, 
$s_0 = 1$ GeV$^2$ and $B(t,Q^2)$ is related to the radius associated with the 
proton vertex.

Models based on the Regge phenomenology have been proposed to describe the  
Deeply Virtual Compton Scattering (DVCS) amplitude 
\cite{CFFJP,Mullerfit,Szczepaniak,Mueller}. DVCS measurements are an important 
source in
extracting information about General Parton Distributions (GPDs). These 
distributions can offer a holographic picture of the nucleon.

The DVCS data collected at the lepton-proton collider HERA can help to
understand the properties of the Pomeron trajectory.  
There are many papers discussing in
details the form and the values of the parameters of the Pomeron
trajectory as well as their possible $Q^2$ dependence (for a recent
review, see, e.g. \cite{Levy}).


In the present paper we compare the model described in Ref.~\cite{CFFJP} and 
applied in
Ref.~\cite{FJ} to the high-energy data on DVCS collected by
the H1 and ZEUS detectors at HERA~\cite{H1_08,H1,ZEUS}. The model makes use 
of a logarithmic Pomeron Trajectory. The scattering amplitude has 
the form~\cite{CFFJP, FJ}

\begin{equation}\label{A2}
A(s,t,Q^2)_{\gamma^* p\rightarrow\gamma p}= -A_0e^{b\alpha(t)}e^{b
\beta(z)}(-is/s_0)^{\alpha(t)}= -A_0e^{(b+L)\alpha(t)+b\beta(z)},
\end{equation}
where $L\equiv\ln(-is/s_0)$, the trajectory at the proton vertex
is
\begin{equation}\label{alpha}
\alpha(t)=\alpha_0-\alpha_1\ln(1-\alpha_2 t),
\end{equation}
whereas the trajectory at the photon vertex is
\begin{equation}\label{beta}
\beta(z)=\alpha_0-\alpha_1\ln(1-\alpha_2 z),
\end{equation}
$z=t-Q^2$ being a new variable introduced in Ref.~\cite{CFFJP}.

For not too
large $Q^2$ the contribution from longitudinal photons to DVCS is small
(it vanishes for $Q^2=0$). Moreover, at the high energies typical
of the HERA collider, the amplitude is dominated by the exchange 
a Pomeron conserving the helicity and, since the final photon is real
and transverse, the initial one is also transverse and the
helicity is conserved. Consider, instead, that electroproduction of vector mesons
requires to take into account both longitudinal and transverse
cross sections. 

The cross section is given by

\begin{equation}{d\sigma\over{dt}}(s,t,Q^2)={\pi\over{s^2}}|A(s,t,Q^2)|^2.
\label{A6}
\end{equation}
and the slope of the forward cone is 

\begin{equation}B={d\over{dt}}\ln(|A(s,t,Q^2)|^2).
\label{A7}
\end{equation}

\begin{figure}[htbp]
\hspace{-1.cm}
\includegraphics[width=0.8\textwidth,angle=0]{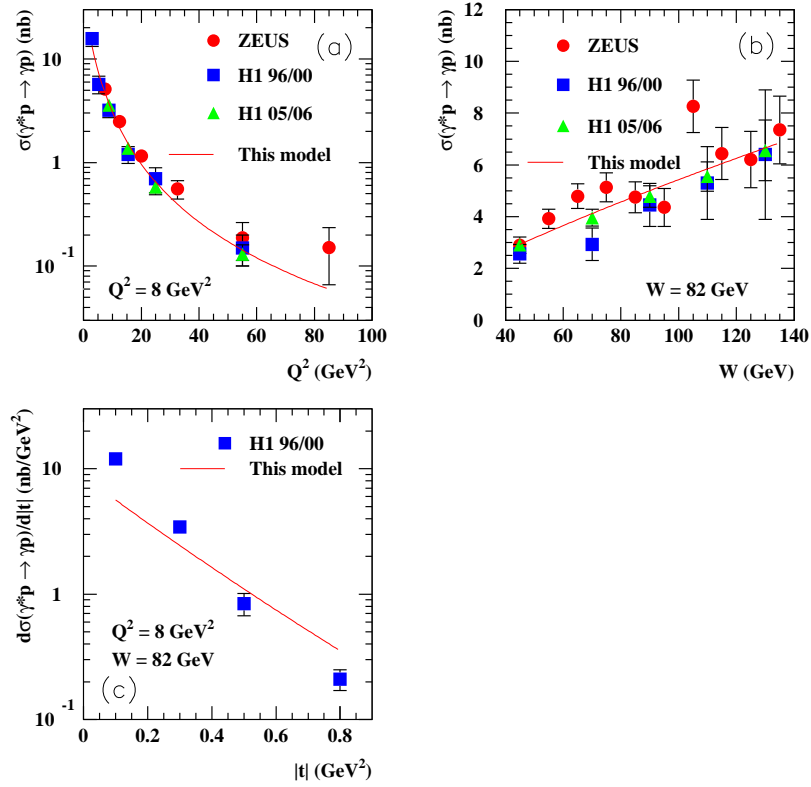}
\caption{Comparison of the model of Ref.~\cite{CFFJP} with the DVCS 
data~\cite{H1_08,H1,ZEUS}. The cross section as a function of
$Q^2$ (a) and $W$ (b), and the differential cross section (c) 
are shown.}
\label{fig:fit_new}
\end{figure}

The parameters $s_0$, $\alpha_1$ and $\alpha_2$ can
be fixed by theoretical constraints, as explained in
Ref.~\cite{CFFJP}. The free parameters are the intercept $\alpha_0$, the 
slope $\alpha$ and the normalization factor $A_0$. Fit to the DVCS 
data collected at HERA~\cite{H1_08,H1,ZEUS} was performed and the 
results are shown in Fig.~\ref{fig:fit_new}.  
The comparison of the model with data for the cross section as a function 
of $Q^2$ and $W$ is shown in 
plot~\ref{fig:fit_new}a and plot~\ref{fig:fit_new}b, 
respectively, while the comparison with the H1 data
for the differential cross section is shown in plot~\ref{fig:fit_new}c . 
The fit suggests that the Pomeron in the DVCS has the intercept $\alpha_0\simeq
1.2$, greater than that in hadronic reactions, and has the slope $\alpha
'\simeq 0.25$, value typical of hadronic scattering. Then the Pomeron 
trajectory, according to this fit, seems to be not so flat as 
it was observed in the diffractive electroproduction of vector
mesons in hard regime.

The fit of $d\sigma/dt$ was performed with all the parameters
fixed excepted for the normalisation factor. One can see from 
plot~\ref{fig:fit_new}c that the model does not agree 
with the H1 measurements (for a relevant discussion see Ref. \cite{FJ}).

\begin{figure}[htbp]
\hspace{-1.cm}
\includegraphics[clip,scale=0.35]{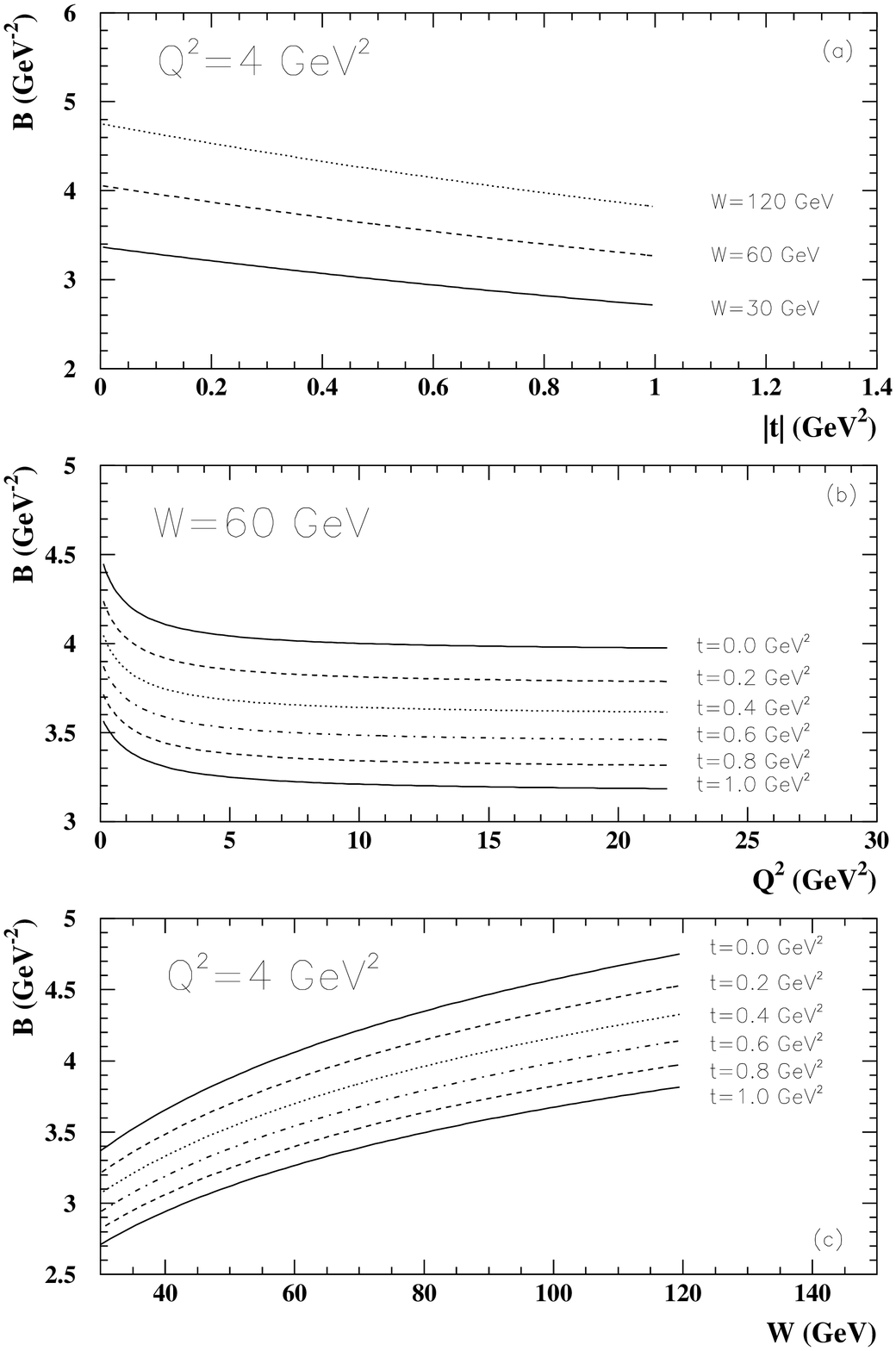}
\caption{Slope behaviour, as calculated from
Eq.~(\ref{A7}),  for several values of $W$, $Q^2$ and $t$.}
\label{fig:slope}
\end{figure}

The slope $B$  (see Eq.~\ref{A7})
is predicted to slightly rise with $W$, but to be almost independent
of $Q^2$.
Fig.~\ref{fig:slope} shows
the slope dependence for different values of $W$ and $tfigut$. 
In particular, the $t$ dependence is
shown in plot~\ref{fig:slope}a at $Q^2=4$ GeV$^2$ for three
different values of $W$. 
The shrinkage of the slope for several values of $t$ 
as a function of $Q^2$ at $W=60$ GeV is shown in plot~\ref{fig:slope}b  
and as a function of $W$ at $Q^2=4$ GeV$^2$ in Fig.~\ref{fig:slope}c.

\vskip0.2cm

\begin{figure}[htbp]
  \includegraphics[height=.35\textheight]{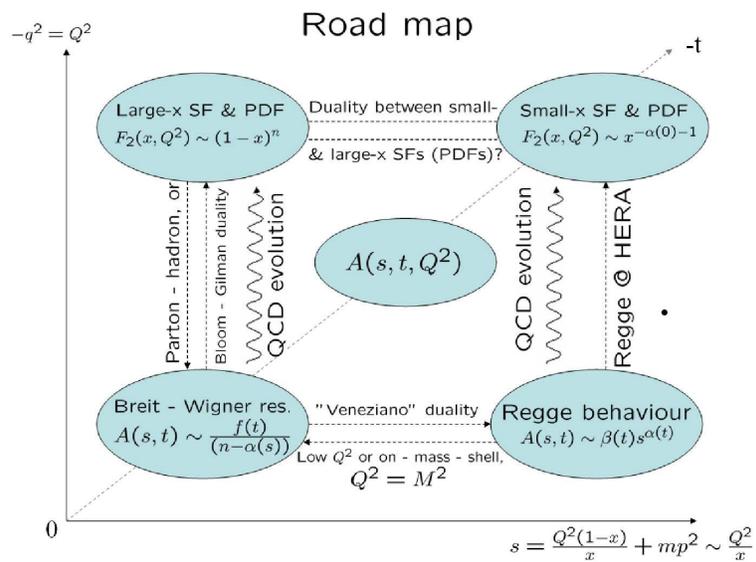}
  \caption{A Road map linking different asymptotic regimes of the
  scattering amplitude $A(s,t,Q^2)$~\cite{Enrico} is shown.}
  \label{fig:roadmap}
\end{figure}

One of the main goals in studying DVCS is the possible extraction
of the GPDs. In view of the difficulties encountered in performing 
the convolution procedure (see, e.g., Ref.~\cite{Szczepaniak, Mueller}), 
the construction of any explicit
model that satisfies the theoretical requirement, and fits the
data, is important, since the  amplitudes of such a model, in the first
approximation, can be identified with the GPDs \cite{Acta}. The
model considered in the present paper is valid in the limit of high
energies. At low energies, where direct channel resonances
dominate, a similar model was constructed in Ref. \cite{Jlab}.
Experimentally, DVCS was and is studied in different kinematical
regions, from low energies (at JLab), through intermediate
energies (by COMPASS) to high energies (at HERA). The
unification of the asymptotic solutions is an ambitious goal of
the theory, and duality may play here an important role, as shown
in the ``Road map'' of Ref.~\cite{Enrico} and in Fig.~\ref{fig:roadmap}. 
In this figure, the present model is described in the upper 
right corner (small $x$ and large $Q^2$, the third variable, $t$, 
being ``compactified''). There is a link between low and high
energies along the horizontal axes, at the bottom of the figure, 
corresponding to the low-$Q^2,$ or the on-mass-shell
connection between the resonance region and the high-energy smoothly 
asymptotic region. This connection was established by means of the dual
models. A similar one, between low and high $x$ values, should
exist also for the off-mass-shell scattering or the structure function 
(upper part of Fig.~\ref{fig:roadmap}).

\begin{theacknowledgments}
We thank the organizers of the the Workshop ``Diffraction 2008'' for their
hospitality. We thank also M. Capua and F. Paccanoni for helpful discussions.
\end{theacknowledgments}

\end{document}